# Improving the mass transfer rate and energy efficiency of solar still by enhancing the inner air circulation

Guilong Peng, Zhenwei Xu, Jiajun Ji, Senshan Sun, Nuo Yang*

State Key Laboratory of Coal Combustion, Huazhong University of Science and Technology, Wuhan 430074, China

*Corresponding email: Nuo Yang (nuo@hust.edu.cn)

**Abstract**

Solar still is an eco-friendly and convenient desalination system that can provide fresh water for remote areas and emergencies. The energy efficiency and productivity of conventional solar still are unsatisfying and need improvement, which requires a deep understanding of the heat and mass transfer process in solar still. In this work, the effect of the inner air circulation on the system's heat and mass transfer performance and energy efficiency are studied theoretically and experimentally. The theoretical results reveal that a weak acceleration of the air circulation inside the SS will significantly increase its performance, due to the improved mass transfer process. By enhancing the inner air circulation, the evaporation and condensation in the solar still can reach up to the limit, and the theoretical energy efficiency reaches up to 87%, 91.5%, and 94.5%, for the input power density at 300 $W/m^2$, 500 $W/m^2$, and 700 $W/m^2$, respectively. Besides, lower ambient temperature and higher ambient convective heat transfer coefficient will decrease the energy efficiency. Given the heat loss, the experimental energy efficiencies are only 3% to 6% lower than the theoretical results, which indicates that the great performance predicted by the theory can be realized in practical application. This work provides a new understanding and strategy for improving the performance of the solar still.

Keywords: Solar still; Desalination; Heat and mass transfer; Convection; Air circulation

## Nomenclature

| | |
|---|---|
| $Cp_{mix}$ | Specific heat capacity of the saturated moist air [J/(kg·K)] |
| $Cp_w$ | Specific heat capacity of the water [J/(kg·K)] |
| $d$ | Moisture content of the dry air [kg/kg] |
| $h_a$ | Convective heat transfer coefficient between the glass cover and the ambient [W/(m$^2$·K)] |
| $h_{h2}$ | Convective heat transfer coefficient of the air circulation [W/(m$^2$·K)] |
| $h_{m2}$ | Convective mass transfer coefficient of the air circulation [m/s] |
| $h_{fg}$ | Latent heat of the vapor [J/kg] |
| $h_{LV}$ | Total enthalpy of phase change [J/kg] |
| $I$ | Input power density [W/m$^2$] |
| $\dot{m}_a$ | Mass flow rate of the dry air [kg/(m$^2$·s)] |
| $\dot{m}_{mix}$ | Mass flow rate of the moist air [kg/(m$^2$·s)] |
| $\dot{m}_v$ | Productivity of the solar still [kg/(m$^2$·s)] |
| $P$ | Total pressure of the air [Pa] |
| $P_f$ | Vapor pressure of the air flow[Pa] |
| $P_F$ | The power of the fan [W] |
| $P_g$ | Vapor pressure at $T_g$ [Pa] |
| $P_v$ | Vapor pressure [Pa] |
| $P_w$ | Vapor pressure at $T_w$ [Pa] |
| $q_{c(g_a)}$ | Heat convection flow between the glass cover and the ambient [J/m$^2$] |
| $q_{c(w_g)}$ | Heat convection flow between the water and the glass cover [J/m$^2$] |
| $q_{d(b_a)}$ | Heat conduction flow between the ambient and the basin [J/m$^2$] |
| $q_{e(w_g)}$ | Heat flow of phase change [J/m$^2$] |
| $q_{r(g_a)}$ | Heat radiation flow between the glass cover and the ambient [J/m$^2$] |
| $q_{r(w_g)}$ | Heat radiation flow between the water and the glass cover [J/m$^2$] |
| $R_h$ | Thermal resistance [(m$^2$·K)/W] |
| $R_m$ | Mass transfer resistance [s/m] |
| $R_{g,v}$ | Gas constant of the vapor [J/(kg·K)] |
| $R_{g,a}$ | Gas constant of the dry air [J/(kg·K)] |
| $SS$ | Solar still |
| $T_{amb}$ | Ambient temperature [K] |
| $T_g$ | Temperature of the glass cover [K] |
| $T_{f1}$ | Temperature of the up-flow moist air [K] |
| $T_{f2}$ | Temperature of the down-flow moist air [K] |
| $T_w$ | Temperature of the water [K] |
| $V$ | Air circulation velocity [m/s] |

Greek letters

| | |
|---|---|
| $\gamma$ | The ratio between the surface area of the glass cover and the water |
| $\varepsilon_1$ | Emissivity between the water and the glass cover |
| $\varepsilon_2$ | Emissivity between the glass cover and the ambient |
| $\varepsilon_w$ | Emissivity of the water |
| $\varepsilon_g$ | Emissivity of the glass cover |
| $\eta_s$ | Energy efficiency of solar still [%] |
| $\rho_v$ | Density of the vapor [kg/m$^3$] |
| $\rho_{mix}$ | Density of the moist air [kg/m$^3$] |
| $\sigma$ | Stefan-Boltzmann constant [W/(m$^2$·K$^4$)] |
| $\emptyset$ | Relative humidity [%] |

## 1. Introduction

The demand for freshwater increases dramatically in the last decades due to the development of agriculture and industry, as well as the growth of population [1]. It is reported that 4 billion people are suffering from freshwater scarcity [2]. Given the abundance of seawater, it is promising to obtain freshwater by desalination [3]. Solar desalination is a popular and eco-friendly desalination technology that draws much attention in both industry and academic fields [4-6]. Solar stills (SS) are the most basic solar desalination systems that are especially suitable for off-grid freshwater supply in remote areas, islands, ships, and so on, for daily or emergent conditions.

Nevertheless, the efficiency and productivity of the conventional SS are low and require further development. During the last decades, solar still was studied from various aspects, such as passive and active working conditions [7, 8], fabricating advanced micro/nano-materials [5, 9, 10], as well as different system constructions [11-13]. Although SS is studied extensively, the energy efficiency of a single-stage SS is usually limited to around 30-60% even with complex modifications [14, 15]. The performance of solar still is limited by many aspects, such as the solar reflectance by the condensation cover, ineffective solar absorption inside the SS, the limited inner mass transfer rate, and so on [16]. Nowadays, with the development of the new system and new materials [17-19], solar reflectance, solar absorption, and water evaporation are no longer the bottlenecks [13, 20, 21]. Therefore, the inefficient mass transfer process is a key point that remains to be investigated and improved.

There are many conventional experimental methods to enhance the mass transfer rate and the energy efficiency of SS. Most of them focus on increasing the temperature difference between the evaporation and condensation surface, such as using film cooling [22], thermal-electrical cooling [23], external condenser [24], heat localization on the evaporation surface [25, 26], and so on. These methods increase the difference of vapor pressure between the evaporation and condensation surface, which increases the mass transfer rate and the efficiency of SS. Meanwhile, a few previous works focus on enhancing the air circulation in SS to improve the performance[27]. However, the understanding of the enhancement by improving the air circulation is very limited.

Besides, several theoretical models were also proposed for analyzing the heat and mass transfer process in SS [28]. Dunkle's model [29] is the most widely used model for analyzing SS, which provides the mass transfer coefficient based on the heat and mass analogy. In Dunkle's model, the mass transfer coefficient only depends on the temperature of the evaporation and condensation surface. The effect of inner air convection cannot be considered, which might lead to a larger error when applying it to different SS. Kumar and Tiwari model [30] is a more realistic model by considering the effects of solar still cavity, operating temperature range, and orientation of condenser cover. Other models like Zheng Hongfei et al.'s model [31] and Tsilingiris model [32] are also well established for understanding the heat and mass transfer process in SS. Nevertheless, it remains unknown that what is the detailed effect of the inner air circulation on the heat and mass transfer, as well as the energy efficiency of SS. Let alone the limit that can be achieved by improving the inner air circulation.

In this work, both the theoretical and experimental analyses for investigating the effect of the inner air circulation on SS are carried out. Firstly, the theoretical model is proposed based on heat and mass conservation. Then, the effects of the input power density, inner air circulation velocity, ambient convective heat transfer coefficient, as well as ambient temperature on the energy efficiency are studied theoretically. The theoretical limit of energy efficiency is given by these analyses. Lastly, experiments are carried out and compared with the theoretical predictions. This work offers a new understanding and strategy for improving the heat and mass transfer, as well as the energy efficiency of SS.

## 2. Methods

### 2.1 Theoretical model

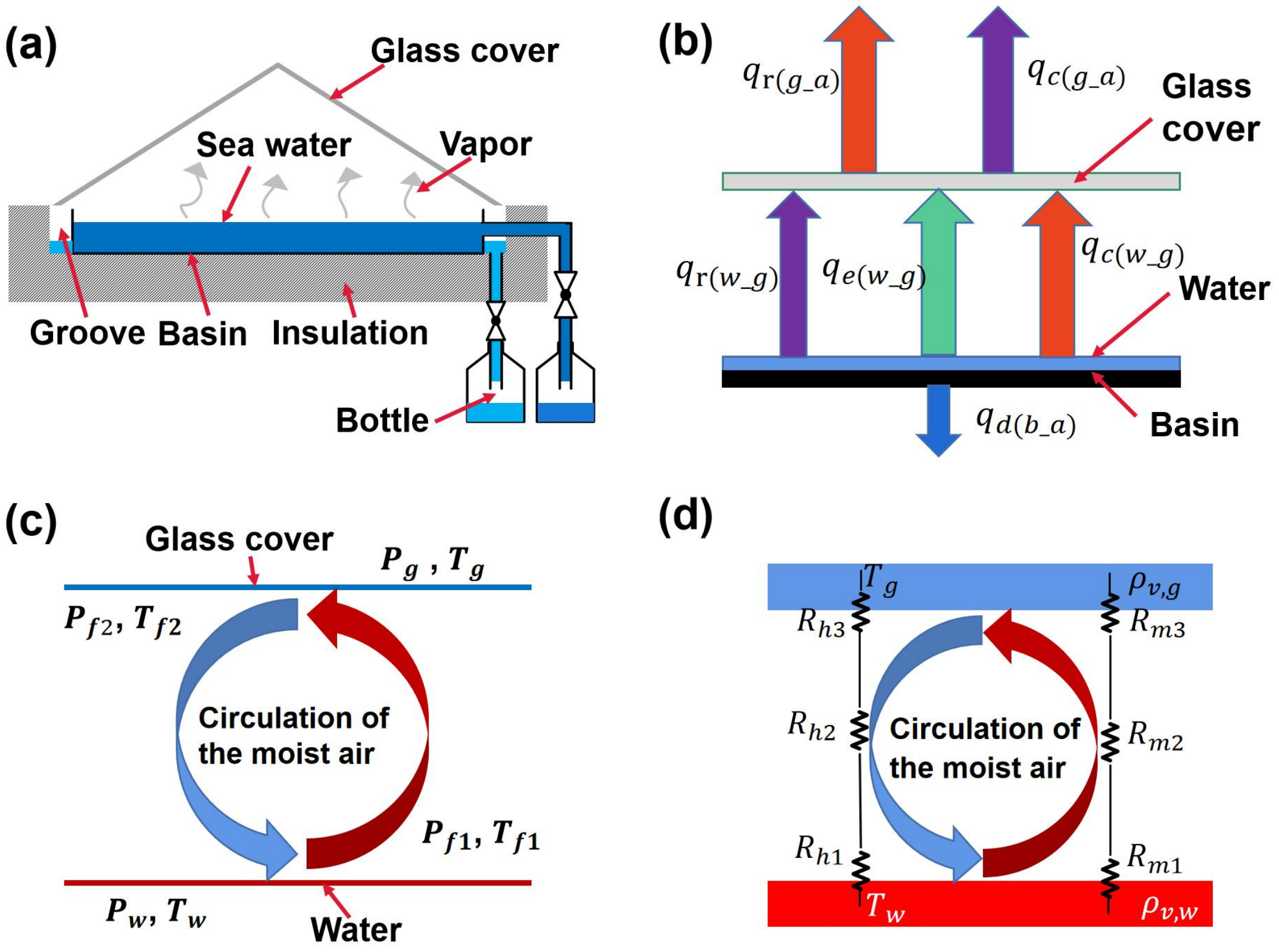

Fig. 1 Theoretical model of the SS. (a) The schematic diagram of a typical SS. (b) The heat transfer process in SS. (c) Model of the ideal convective air circulation in SS. (d) Model of the heat and mass transfer resistance in SS.

A typical solar still consists of six main parts as shown in Fig. 1a, including the basin, seawater, glass cover, insulation, grooves, and bottles. The basin absorbs solar irradiation and heats the seawater above it. Then, the heated seawater evaporates and the vapor rises to the glass cover. The temperature of the glass cover is lower than that of water, thus the vapor condenses as droplets on the glass cover. Later, the droplets grow up and finally flow down as freshwater to the grooves that are connected with a bottle. The high concentration brine will be rejected to another bottle. To decrease the heat loss, the insulation material is used on the basin and walls.

The main heat transfer process of SS is shown in Fig. 1b. The absorbed energy in the basin is either transported to the above water or dissipated to the ambient. The energy in the water is further transported to the glass cover by the air convection, radiation, and phase change processes. The temperature of the glass cover remains relatively low by dissipating the heat to the ambient via heat convection and radiation. Based on the energy conservation, it can be obtained that,

$$I = q_{r(w_g)} + q_{d(b_a)} + q_{e(w_g)} + q_{c(w_g)} \qquad (1)$$

where $I$ is the input power density, corresponding to the absorbed solar energy in practical applications. $q_{r(w_g)}$, $q_{e(w_g)}$, $q_{c(w_g)}$ are the heat flows from the water to the glass cover by radiation, phase change, and convection, respectively. $q_{d(b_a)}$ is the heat flow from the basin to the insulation and ambient by heat conduction, which is regarded as zero in the ideal model. The overall energy efficiency, $\eta_s$, of SS is,

$$\eta_s = \frac{q_{e(w_g)}}{I} \times 100\% \qquad (2)$$

In SS, the mass transfer is mainly driven by air convection as shown in Fig. 1c. The air is heated and humidified near the water surface to be saturated at temperature $T_{f1}$, and vapor pressure $P_{f1}$. Then the moist air is cooled and dehumidified by the glass cover to be $T_{f2}$ and $P_{f2}$. For ideal air circulation, the temperature and vapor pressure of air only change at the interface, where heat and mass transfer happen.

The air convection in a conventional SS is driven by buoyancy, thus very weak and leads to large thermal and mass transfer resistance. The resistance can be divided into 3 parts as shown in Fig. 1d, including the interface resistance between water and air $R_1$, air and glass $R_3$, as well as the resistance of air convection $R_2$. The subscripts $h$ and $m$ indicate the thermal and mass transfer, respectively. On one hand, the weak natural convection in SS enables a low convective heat loss from the water to the glass cover. On the other hand, the vapor transport is also suppressed, thus the phase change process is also limited. Therefore, it is questionable that what the efficiency will be when increasing the air convection.

Herein, to investigate the effect of inner air convection on the heat and mass transfer, an ideal model is proposed by making the following assumptions:

a) The temperature of the air is heated to the water temperature $T_w$ before leaving the water-air interface, and cooled down to the glass temperature $T_g$ before leaving the air-glass interface, i.e.,

$$T_{f1} = T_w \tag{3}$$

$$T_{f2} = T_g \tag{4}$$

b) The air temperature remains constant and the humidity remains saturated, after leaving the water-air interface and before reaching the air-glass interface.

c) The temperature as well as the heat and mass transfer rates are uniform at the interfaces.

d) The heat loss from the basin to the ambient is zero.

Based on the assumptions, the heat flow of phase change can be obtained by,

$$q_{e(w_g)} = h_{LV}\dot{m}_v \tag{5}$$

where $h_{LV}$ is the enthalpy of phase change, which equals to the summation of the latent heat $h_{fg}$ and the sensible heat [33].

$$h_{LV} = h_{fg} + Cp_w(T_w - T_{amb}) \tag{6}$$

$$h_{fg} = 1.91846 \times 10^6[\frac{T_w}{T_w - 33.91}]^2 \tag{7}$$

where $Cp_w$ is the thermal capacity of the moist air at $T_w$. $T_{amb}$ is the temperature of the ambient.

$\dot{m}_v$ is the productivity of the fresh water in SS, which is related to the mass flow rate of the dry air $\dot{m}_a$ and the moisture content $d$,

$$\dot{m}_v = \dot{m}_a(d_w - d_g) \tag{8}$$

where $d_w$ and $d_g$ are the moisture content of the dry air at $T_w$ and $T_g$, respectively, which can be calculated based on the vapor pressure $P_v$,

$$d = \frac{R_{g,a}}{R_{g,v}}\frac{P_v}{P - P_v} \tag{9}$$

where $R_{g,a} = 287\,J/(Kg \cdot K)$ and $R_{g,v} = 461\,J/(Kg \cdot K)$ are the gas constant of dry air and water vapor, respectively. $P$ is the total pressure of the moist air, which is regarded as 101 kPa in the model. The vapor pressure $P_v$ can be obtained according to the vapor temperature $T_v$ and the relative humidity $\emptyset$ [28],

$$P_v = \emptyset e^{\left(25.317 - \frac{5144}{T_v}\right)} \quad (10)$$

$\dot{m}_a$ is related to the total mass flow rate of the moist air $\dot{m}_{mix}$,

$$\dot{m}_a = \frac{\dot{m}_{mix}}{(1 + d_w)} \quad (11)$$

where $\dot{m}_{mix}$ is decided by the air circulation velocity $V$ and the moist air density $\rho_{mix}$,

$$\dot{m}_{mix} = \frac{1}{2} V \rho_{mix} \quad (12)$$

Herein, the constant coefficient '1/2' indicates that half of the basin area is occupied by the hot air that flows upward, and the other half is occupied by the cold air that flows downward.

Table 1 List of coefficients for calculating the thermos-physical property of the saturated moist air (0 – 100 °C).

| | 0 | 1 | 2 | 3 | 4 | 5 |
|---|---|---|---|---|---|---|
| SD | 1.293 | -5.538E-3 | 3.860E-5 | -5.254E-7 | – | – |
| SC | 1.005 | 2.051E-3 | -1.632E-4 | 6.212E-6 | -8.83E-8 | 5.071E-10 |

The density $\rho$ and the heat capacity *Cp* of the saturated moist air at temperature *t* (in Celsius degree) can be obtained by the following equations and coefficients in Table 1 [34],

$$\rho_{mix,t} = \mathrm{SD}_0 + \mathrm{SD}_1 t + \mathrm{SD}_2 t^2 + \mathrm{SD}_3 t^3 \quad (13)$$

$$Cp_{mix,t} = \mathrm{SC}_0 + \mathrm{SC}_1 t + \mathrm{SC}_2 t^2 + \mathrm{SC}_3 t^3 + \mathrm{SC}_4 t^4 + \mathrm{SC}_5 t^5 \quad (14)$$

The convective heat flow rate $q_{c(w_g)}$ can be calculated by,

$$q_{c(w_g)} = \dot{m}_{mix} Cp_w \left(T_w - T_g\right) \quad (15)$$

The irradiation heat flow rate $q_{r(w_g)}$ is,

$$q_{r(w_g)} = \varepsilon_1 \sigma ({T_w}^4 - {T_g}^4) \quad (16)$$

where $\sigma = 5.67 \times 10^{-8}\ \mathrm{W/(m^2 \cdot K^4)}$ is the Stefan-Boltzmann constant, $\varepsilon_1$ is the emissivity between the water and glass,

$$\varepsilon_1 = \left[\frac{1}{\varepsilon_w} + \frac{1}{\varepsilon_g} - 1\right]^{-1} \tag{17}$$

where $\varepsilon_w$ and $\varepsilon_g$ are the emissivity of the water and glass, respectively.

The energy conservation on the glass cover is,

$$q_{r(w_g)} + q_{e(w_g)} + q_{c(w_g)} = \gamma q_{r(g_a)} + \gamma q_{c(g_a)} \tag{18}$$

where $\gamma$ is the ratio between the glass cover area $A_g$ and water area $A_w$,

$$\gamma = \frac{A_g}{A_w} \tag{19}$$

The convective heat transfer rate between the glass cover and the ambient $q_{c(g_a)}$ is,

$$q_{c(g_a)} = h_a\left(T_g - T_{amb}\right) \tag{20}$$

where $h_a$ is the convective heat transfer coefficient of the ambient air.

The radiative heat transfer rate between the glass cover and the ambient $q_{r(g_a)}$ is,

$$q_{r(g_a)} = \varepsilon_2\sigma\left({T_g}^4 - {T_{amb}}^4\right) \tag{21}$$

where $\varepsilon_2$ is the emissivity between the glass and the ambient.

The convective heat transfer coefficient $h_{h2}$ and the mass transfer coefficient $h_{m2}$ of the air circulation in SS are，

$$h_{h2} = \frac{q_{c(w_g)}}{T_w - T_g} \tag{22}$$

Therefore, $h_{h2}$ can be rewritten as

$$h_{h2} = \dot{m}_{mix} Cp_w \tag{23}$$

$$h_{m2} = \frac{\dot{m}_v}{\rho_{v,w} - \rho_{v,g}} \tag{24}$$

The vapor density $\rho_v$ is also decided by the vapor temperature $T_v$ as,

$$\rho_v = \frac{P_v}{R_{g,v}T_v} \tag{25}$$

Thus, the heat and mass transfer coefficients, water and glass temperatures, as well as energy efficiency can be obtained by solving the above equations. The efficiency $\eta_s$ can be rewritten as

$$\eta_s = \frac{h_{LV} V_{mix} \rho_{mix}}{2I} \frac{(d_w - d_g)}{(1 + d_w)} \quad (26)$$

2.2 Experimental setup

To verify the practicality of theoretical predictions, experiments are carried out by an indoor experimental setup. The photo of the setup is shown in Fig. 2a. Firstly, a pyramid SS with the size of 25 cm × 25 cm is fabricated. A heating panel is placed under the basin for heating the basin and water as shown in Fig. 2b. 5 mm thickness of water is placed in the basin to be treated. The power of the heating panel is controlled by a power supply. The basin of the solar still is insulated by 4 cm thickness of XPS (extruded polystyrene) foam, and the angle of the glass cover is fixed at 30°.

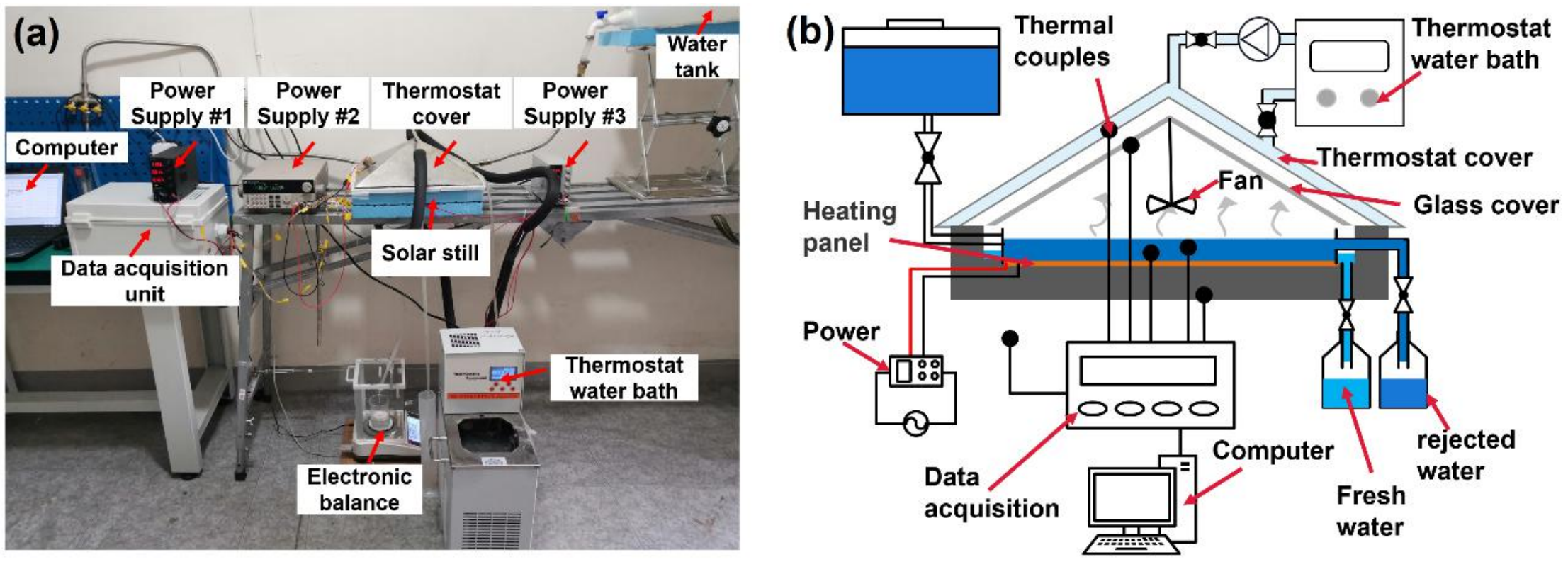

Fig. 2 (a) Photo of the experimental setup. (b) Schematic diagram of the experimental setup.

A thermostat cover is placed above the glass cover to simulate and control the ambient temperature. The air gap between the glass cover and the thermostat cover is 1 cm. The thermostat cover has the same shape as the glass cover and its temperature is controlled by circulating the water from a thermostat water bath. The validation of the thermostat cover is shown in Supporting Information Note S1. The freshwater is collected by a beaker on the electronic balance. The temperatures are measured by K type of thermal couples and the temperature data is collected by a data acquisition unit.

The installation detail of thermal couples is shown in Fig. S2. All the specifications of the devices and sensors are listed in Table 2.

A fan with a diameter of 5 cm is installed inside of the SS for enhancing the inner air circulation velocity (Fig. 2b). The fan is fixed above the center point of the basin. The distance between the fan and the water surface is 1 cm. During the experiments, the fan actively transports the moist air near the water surface to the glass cover. To simulate different air circulation velocities, a power supply is used to control the power of the fan.

## 2.3 Uncertainty analysis

Table 2 Specific of devices and sensors in the experiments.

| Name | Brand | Type | Function | Range | Error |
|---|---|---|---|---|---|
| Fan | LFFAN | LFS0512SL | Enhancing air circulation | 0 ~ 4800 RPM | 10% RPM |
| Electronic balance | ANHENG | AH-A503 | Measuring productivity | 0 ~ 500 g | ± 0.01 g |
| Power supply #1 & #3 | WANPTEK | NPS3010W | Supplying power for the data acquisition unit and heating panel | 0 ~ 30 V | ± 0.1 % |
| Power supply #2 | ITECH | IT6932A | Supplying power for the fan | 0 ~ 60 V | ± 0.03 % |
| Data acquisition unit | CAMPBELL SCIENTIFIC | CR1000X | Collecting and saving data | - | ± 0.01 °C |
| Thermal couple multiplexer | CAMPBELL SCIENTIFIC | AM25T | Transmitting temperature data | 0 ~ 25 Channels | - |
| Thermostat water bath | QIWEI | DHC-2005-A | Controlling the ambient temperature around the glass cover | -20 ~ 99.9 °C | ± 0.2 °C |
| Heating panel | BEISITE | Custom-made | Heating the water | 0 ~ 2000 W/m$^2$ | - |
| Thermal couple | ETA | T-K-36-SLE | Measuring the temperature | -200 ~ 260 °C | ± 1.1 °C |

The diameter and thickness of the fan in this work are 5 cm and 1 cm, respectively. The nominal voltage, current, rotate speed and air volume of the fan are 12 V, 0.1 A, 4800 RPM (Revolutions Per Minute), and 10.5 CFM (Cubic Feet Per

Minute), respectively. The working voltage is controlled to provide different powers of the fan, which simulate different air circulation velocities. The mass change of the freshwater was measured by the electronic balance every 10 seconds for calculating the productivity. The temperature data were also collected and saved by the data acquisition unit every 10 seconds. The specifications and functions of the devices and sensors are listed in Table 2.

## 3. Results and discussion

### 3.1 Theoretical results

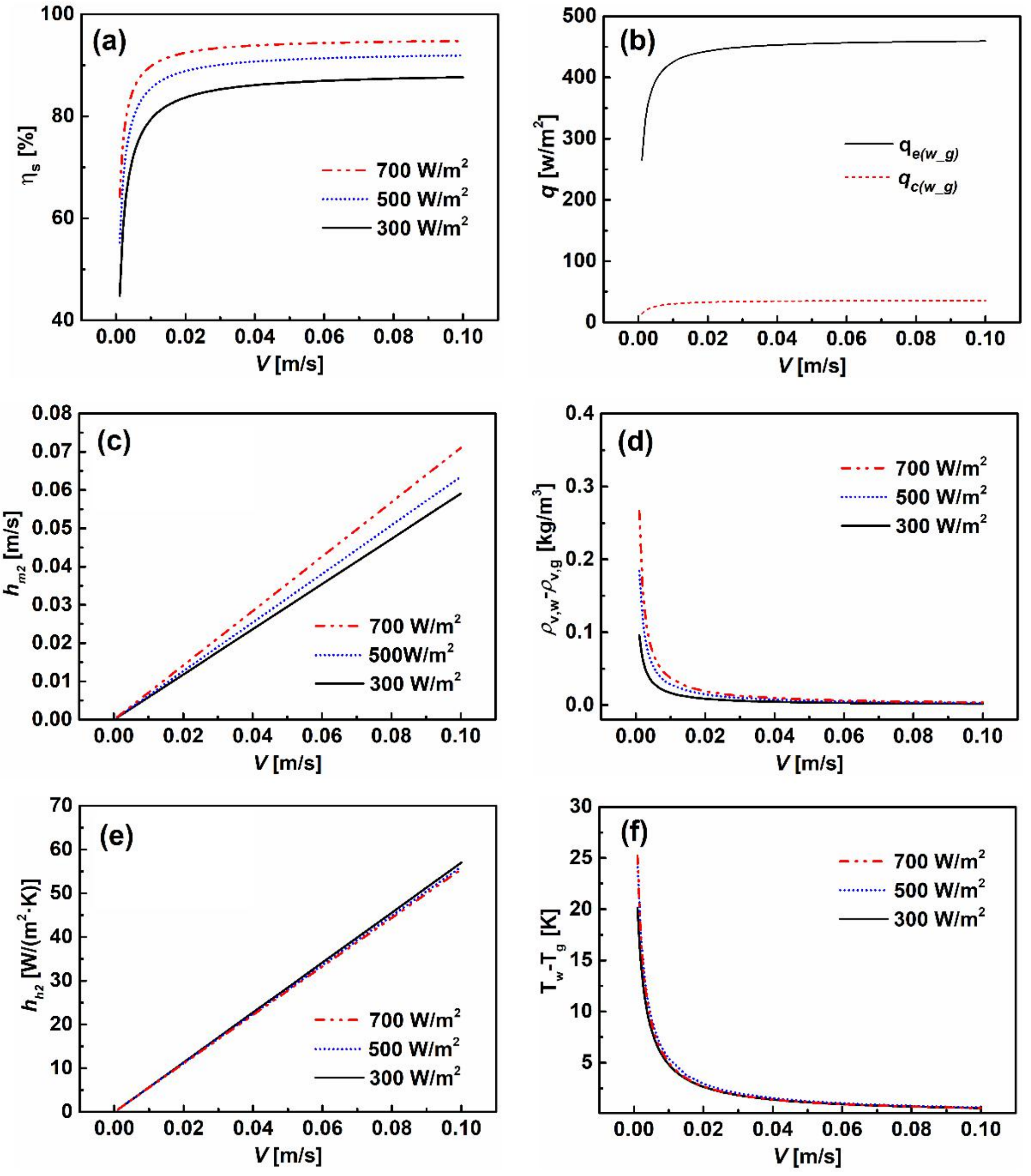

Fig. 3 Theoretical results of SS, $T_{amb}$, $h_a$, and $\gamma$ are fixed at 30 °C, 10 W/($m^2$·K), and 1.15, respectively. The minimum $V$ in the calculation is 0.001 m/s. (a) The theoretical

efficiency of SS under different input power densities. (b) The heat flow rate of phase change and convection under 500 W/$m^2$ of input power density. (c) The mass transfer coefficient under different air circulation velocities. (d) The difference in vapor density between the water and the glass cover. (e) The heat transfer coefficient under different air circulation velocities. (f) The difference in temperature between the water and the glass cover.

The theoretical results of SS under different input power densities are shown in Fig. 3. The efficiency increases dramatically before the air circulation velocity is lower than 0.02 m/s, then the efficiency increases very slightly and converges with the further increases of air circulation velocity (Fig. 3a). The results imply that a weak acceleration of the air circulation inside the SS will increase its performance a lot. The enhancement of energy efficiency is attributed to the significant enhancement of the phase change process as shown in Fig. 3b. Both the heat flow rates of phase change and convection increase with the increases of air velocity at first. However, the convective heat flow rate remains very low, thus has a negligible effect on the overall efficiency.

At a low air circulation velocity, the heat flow rate by phase change is relatively low, which indicates a limited mass transfer process (Fig. 3b). Thereby, a significant part of heat will be dissipated and wasted by thermal radiation, convection, and conduction. It should be noted that the ideal model assumes sufficient heat and mass transfer rates at the interfaces for all the air circulation velocities. However, the heat and mass transfer rates will be insufficient in practical application, especially under low air velocity. Therefore, a higher air circulation velocity will be needed in practical application to approach more sufficient heat and mass transfer rates and high energy efficiency. Meanwhile, it is also found that the difference of vapor density and temperature between the water and glass cover decreases very fast with the increase of air circulation velocity (Fig. 3d and 3f). As a result, energy efficiency converges quickly.

The maximum efficiency of SS increases with the increase of input power density. The maximum efficiency is around 87%, 91.5% and 94.5%, for input power density at 300 W/m$^2$, 500 W/m$^2$ and 700 W/m$^2$, respectively. This is because, on one hand, both the mass transfer coefficient and the difference of vapor density ($\rho_{v,w} - \rho_{v,g}$) increase with the input power density (Fig. 3c and 3d), which indicates an improved mass transfer process. On the other hand, the heat transfer coefficient and the difference of temperature ($T_w$ - $T_g$) are almost the same under different input power densities (Fig. 3e and 3f). Therefore, the heat loss from the water to the glass cover by air convection and thermal radiation is nearly constant. The enhanced mass transfer rate and the constant heat loss enable improved energy efficiency under high input power density.

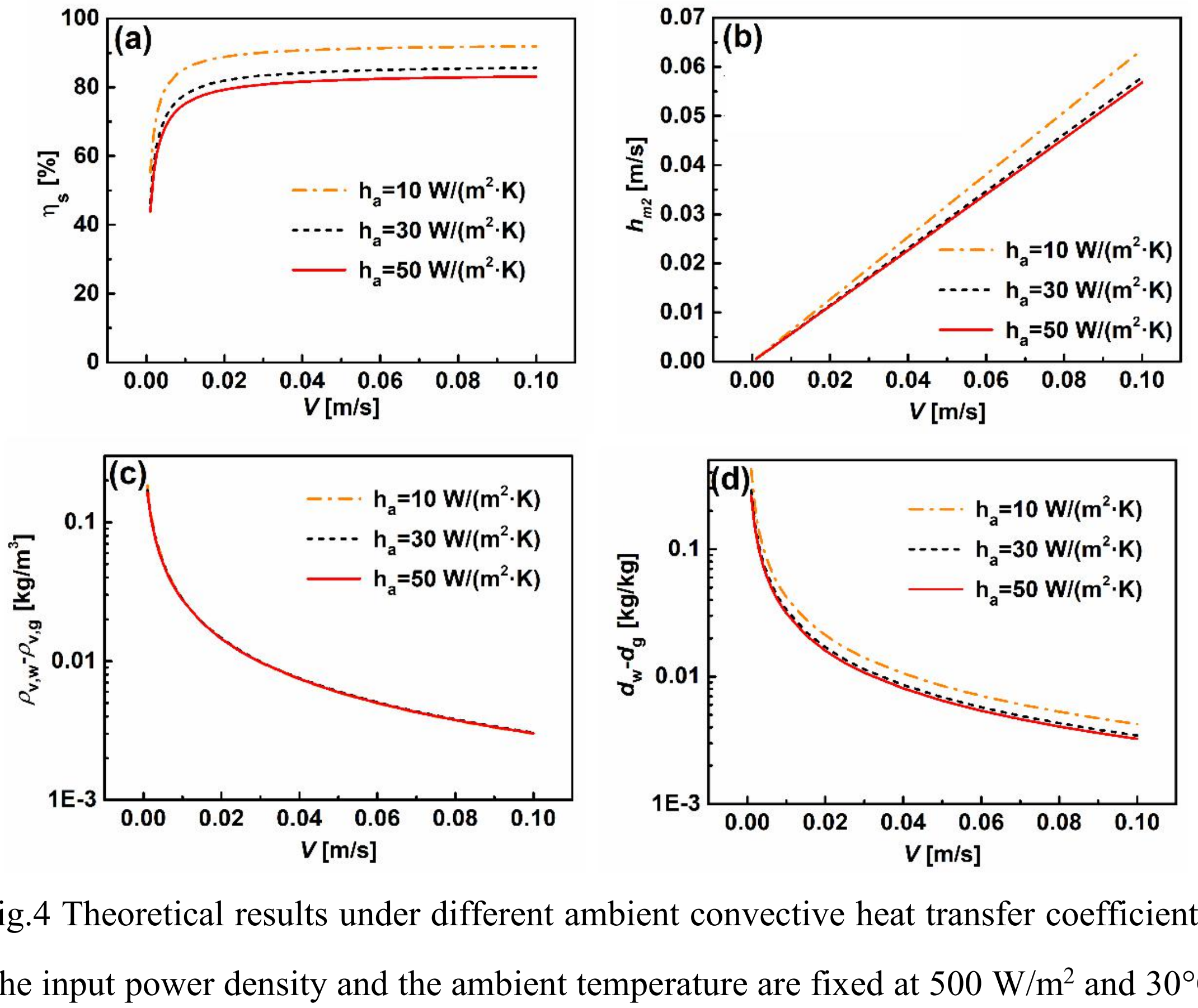

Fig.4 Theoretical results under different ambient convective heat transfer coefficients. The input power density and the ambient temperature are fixed at 500 W/m$^2$ and 30°C, respectively. (a) The energy efficiency of SS. (b) The mass transfer coefficient of inner air circulation. (c) The difference in vapor density between the water and glass cover. (d) The difference in the moist content between the water and the glass cover.

The effect of the ambient convective heat transfer coefficient ($h_a$) is investigated as shown in Fig. 4. In practical application, $h_a$ is affected by the airflow near the glass cover. Herein, a typical natural convection coefficient $h_a$=10 W/(m$^2$·K) is choosing as the coefficient that represents a windless ambient of SS. On the other hand, higher $h_a$ represents a windy ambient. It is shown that a lower $h_a$ gives a higher efficiency (Fig. 4a). The maximum efficiency is around 91.5%, 85% and 82.5%, for $h_a$ at 10 W/(m$^2$·K), 30 W/(m$^2$·K) and 50 W/(m$^2$·K), respectively. For a higher $h_a$, $h_{m2}$ decreases (Fig. 4b), and the difference of vapor density between the water and glass cover is almost the same under different $h_a$ (Fig. 4c). Based on Eq. (8), the mass transfer coefficient is affected by the moist content $d$. A higher difference of moist content $(d_w - d_g)$ indicates more condensable vapor mass for a given mass flow rate of the moist air. It can be found that $d_w - d_g$ decreases with the increases of $h_a$ as shown in Fig. 4d. Therefore, mass transfer coefficient, productivity, and energy efficiency also decrease.

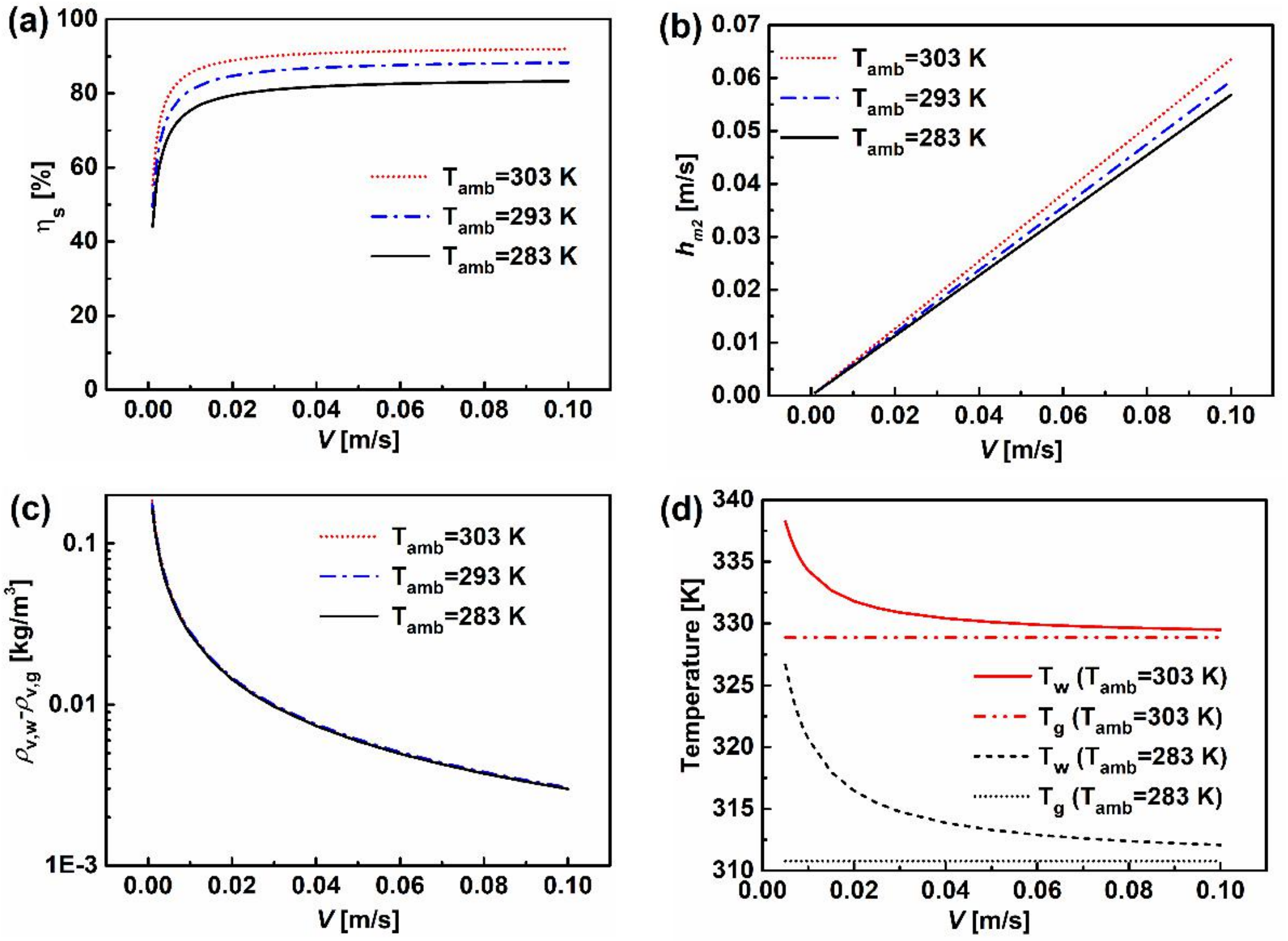

Fig. 5 Theoretical results under different ambient temperatures. $I$ and $h_a$ are fixed at 500 W/m$^2$ and 10 W/(m$^2$·K), respectively. (a) The energy efficiency under different ambient temperatures. (b) The mass transfer coefficient of inner air circulation. (c) The difference in vapor density between the water and glass cover. (d) The temperature of the water and the glass cover under different ambient temperatures.

Moreover, it is found that the ambient temperature also significantly affects the performance of SS (Fig. 5). The energy efficiency decreases with the decreases of the ambient temperature as shown in Fig. 5a. The maximum efficiency is around 91.5%, 88%, and 83%, for $T_{amb}$ at 303K, 293K, and 283K, respectively. Similar to the effect of $h_a$, the ambient temperature affects the mass transfer coefficient instead of the difference of vapor density (Fig. 5b and 5c). The mass transfer coefficient will be higher under a higher ambient temperature. A lower ambient temperature will lead to a higher temperature difference, but lower water and glass cover temperatures as shown in Fig. 5d. The results indicate that, under the same solar insolation, the energy efficiency of SS at hot regions or seasons will be higher than that of cold regions or seasons.

The experiments are further carried out to verify the practicality of theoretical prediction (Fig. 6). In the experiments, a fan fixed in the SS is used to provide different air circulation velocities by operating it under different powers. The energy efficiency of the SS without turning on the fan is 53.4%, 64.8%, and 71.1% at 300 W/m$^2$, 500 W/m$^2$, and 700 W/m$^2$ of input power density, respectively. The measured energy efficiency is higher than that of the outdoor experiments in previous works. This is because that the indoor experiments eliminated the energy loss of light reflections by the glass cover and the basin. Meanwhile, the indoor experiments are carried out under a steady-state instead of the transient state as in outdoor experiments, which eliminated the energy loss of heating the system. Therefore, the indoor experiments represent the best case that the outdoor experiments can achieve, hence better performance.

It is shown that the productivity and energy efficiency of SS increases significantly when the power of the fan ($P_F$) slightly increases from 0 to 0.2 W (Fig. 6a and 6b). Meanwhile, higher input power density also enables higher productivity and energy efficiency as predicted by the theory. The energy efficiency increases to 59.7%, 74.7%, and 78.7% by using 0.075 W of fan power. Then, it is increased to 68.4%, 78%, and 86.2% by using 0.2 W of fan power. The results imply that a slight enhancement of the air circulation will increase the system performance a lot, which agrees with the theoretical prediction very well. At higher $P_F$, the productivity and the energy efficiency converge, which also agree with the theoretical prediction. The maximum enhancement of energy efficiency is around 42%, 28%, and 20% at 300 $W/m^2$, 500 $W/m^{2,}$ and 700 $W/m^2$ of input power density, respectively.

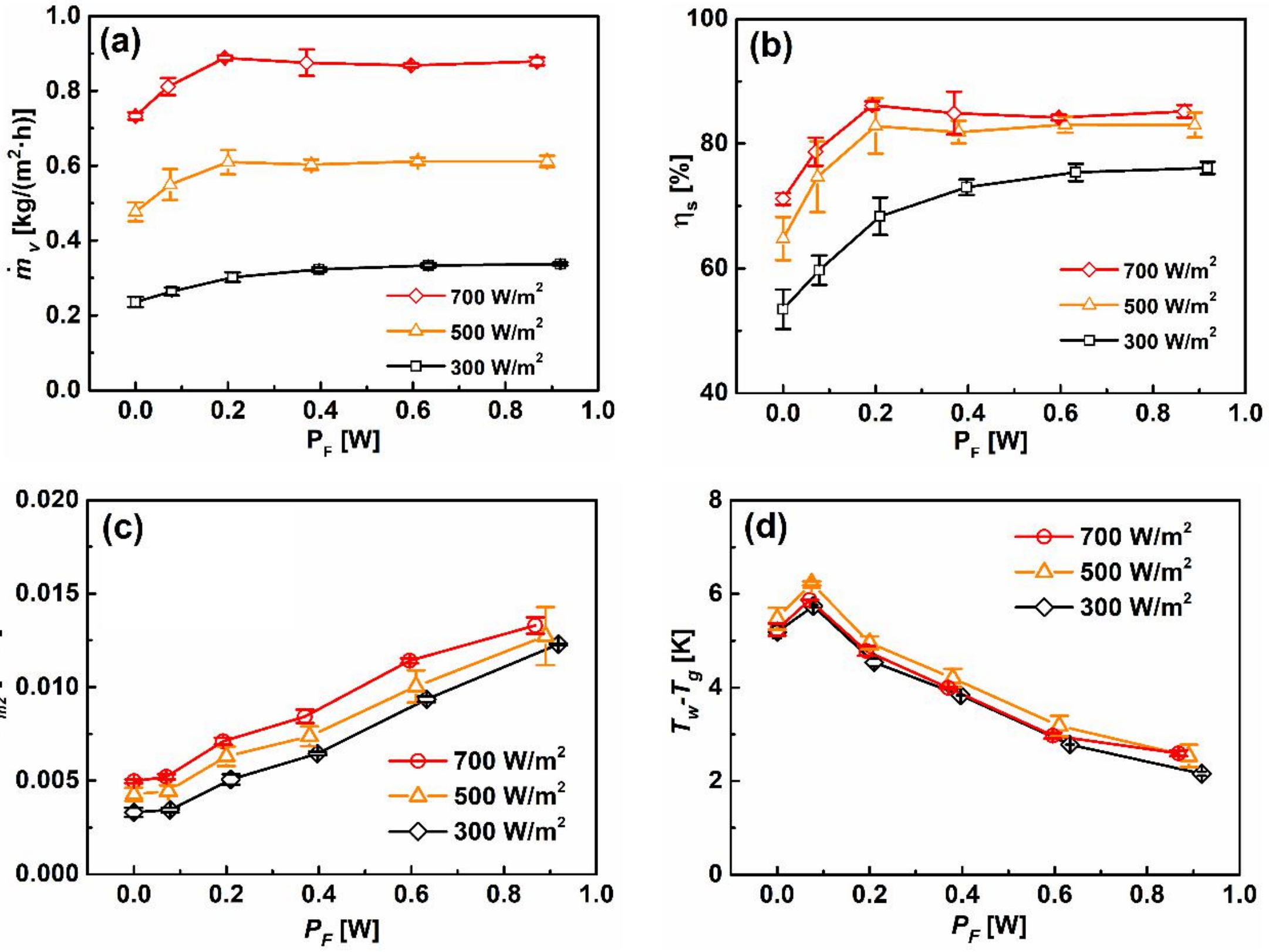

Fig. 6 Experimental results of SS under different power of the fan ($P_F$) and different input power density. The ambient temperature is fixed at 30 °C. (a) The productivity of SS. (b) The energy efficiency of SS. (c) The mass transfer coefficient of the inner air circulation. (d) The temperature difference between the water and the glass cover.

The experimental mass transfer coefficients also increase with the increases of $P_F$. Under natural conditions without turning on the fan, the mass transfer coefficients are lower than 0.005 m/s at the three measured input power densities (Fig. 6c). It indicates that the air circulation is weak under natural convection. The mass transfer coefficients keep increasing with $P_F$ and reach up to more than 0.012 m/s when $P_F$ is around 0.9 W. On the contrary, the temperature difference between the water and the glass cover decreases under high $P_F$ (Fig. 6d). Besides, it is found that the temperature difference is almost the same under different input power densities, which confirms the theoretical prediction. Nevertheless, there is a slight increase of temperature difference when increasing $P_F$ from 0 W to 0.075 W. This might be because that the fan concentrates the hot air at the center area of the glass cover, where most of the vapor condensates. Thus the glass temperature at the side area decreases a lot due to less heat and mass transfer. As a result, the temperature difference increases at first (Supporting information Note S2).

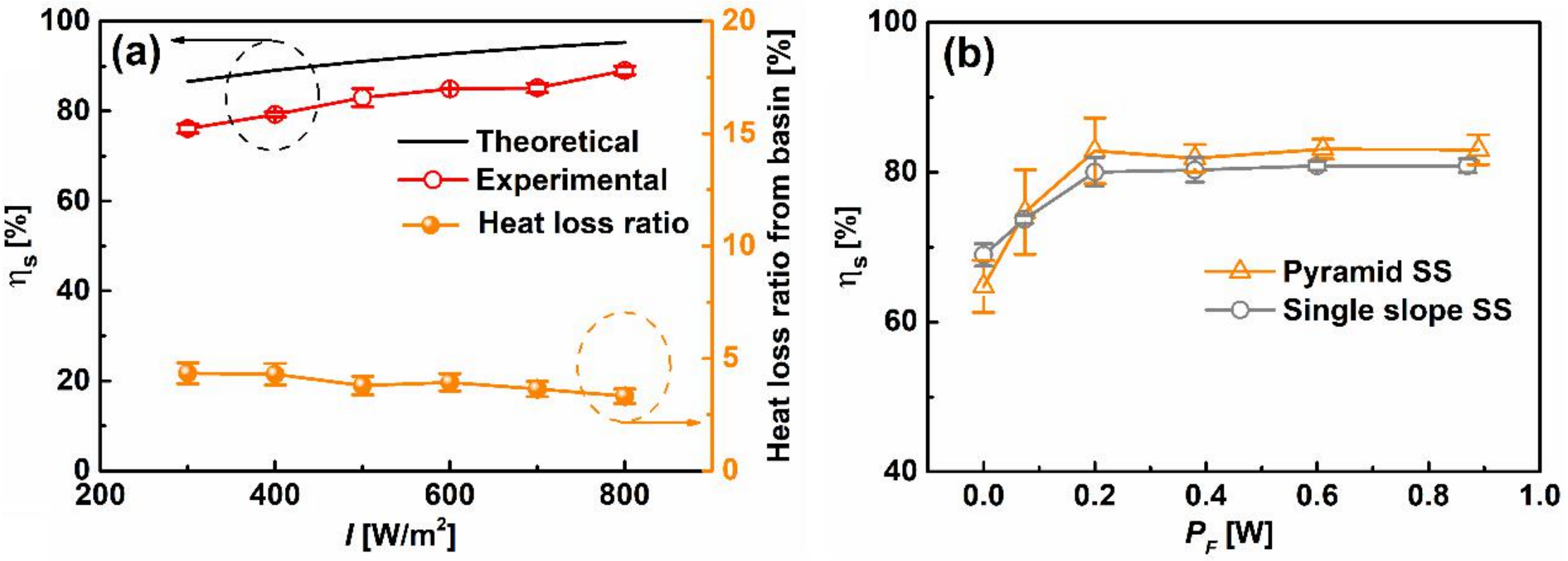

Fig. 7 Comparison of the theoretical and experimental results. (a) Theoretical and experimental results under different input power densities *I*. (b) Experimental energy efficiency of pyramid SS and single slope SS. The input power density is fixed at 500W/m$^2$.

The comparison between the theoretical and experimental maximum energy efficiency is shown in Fig.7. The theoretical maximum energy efficiency of SS increases from 87% to 95.7% with the increases of the input power density from 300 $W/m^2$ to 800 $W/m^2$ (Fig.7a). As a comparison, the maximum experimental energy efficiency increases from 76% to 89%, which is 6% to 11% lower than the theoretical results. One of the main reasons for the difference is the heat loss in the experiments. Based on the temperatures, it is obtained that the heat loss from the basin accounts for around 3% - 5% of the total input power density. Given the heat loss, the experimental energy efficiencies are only 3% to 6% lower than the theoretical predictions, which shows a good agreement between the theoretical and experimental results. Therefore, it can be concluded that the limit predicted by the theory can be reached in practical application, which imply that the evaporation and condensation rate in the solar still is enhanced to near the limit without the need of extra components, such as the external condenser.

In addition, the effects of the inner air circulation on two different types of SS are compared (Fig. 7b). It shows that a low $P_F$ can enhance the energy efficiency of both single slope SS and the pyramid SS. The trend is very similar for the two kinds of SS, which shows the universality of enhancing energy efficiency by improving the inner air circulation. Under natural conditions, the energy efficiency of the single slope SS is slightly higher than that of the pyramid SS. However, the energy efficiency of the single slope SS is lower than that of the pyramid SS when the $P_F$ is higher than 0.2 W, which might be due to the larger heat dissipation area in the single slope SS. The experimental results under different ambient temperatures also agree well with the theoretical results (Fig. S3). The energy efficiency increases slightly with the increases of the ambient temperature in both experiment and theory.

## 4. Conclusion

In conclusion, the inner air circulation affects the performance of the solar still significantly. The theoretical results show that the energy efficiency increases dramatically when the ideal air circulation velocity is less than 0.02 m/s. It is due to that a higher air circulation velocity increases the convective mass transfer rate a lot. On the other hand, the convective heat transfer remains relatively weak, which enables a low heat loss from the water to the glass cover. The maximum theoretical energy efficiency increases from 87% to 95.7% with the increases of the input power density from 300 $W/m^2$ to 800 $W/m^2$. It is also observed that the limit of the mass transfer rate is low at a low air circulation velocity, which limits the energy efficiency of the conventional solar still. Besides, the energy efficiency decreases 9% and 8.5%, when the ambient convective heat transfer coefficient increases from 10 $W/(m^2 \cdot K)$ to 50 $W/(m^2 \cdot K)$ and the ambient temperature decreases from 30 °C to 10°C, respectively.

Meanwhile, the maximum experimental energy efficiency is 6% to 11% lower than the theoretical prediction. One important reason for the difference is the heat loss from the basin to the ambient in experiments, which accounts for 3% - 5% of the total input power density. Moreover, it is found that the inner air circulation has a similar effect on different types of solar still. In general, the experimental results agree well with the theoretical predictions. It can be concluded that the idea performance predicted by the theory is achievable in practical application. The high performance also implies that the evaporation and condensation in solar still can reach up to the limit without the need for external components.

**Conflicts of interest**

There is no conflict of interest to declare.

**Acknowledgment**

The work was sponsored by the National Key Research and Development Program of China (2018YFE0127800), China Postdoctoral Science Foundation (2020M682411), National Natural Science Foundation of China (51950410592) and Fundamental Research Funds for the Central Universities (2019kfyRCPY045), and Program for HUST Academic Frontier Youth Team. The authors thank the National Supercomputing Center in Tianjin (NSCC-TJ) and China Scientific Computing Grid (ScGrid) for assisting in computations.

## References

[1] Rosa L, Chiarelli DD, Rulli MC, Dell'Angelo J, D'Odorico P. Global agricultural economic water scarcity. Sci. Adv. 2020;6:eaaz6031.

[2] Mekonnen MM, Hoekstra AY. Four billion people facing severe water scarcity. Sci. Adv. 2016;2 e1500323.

[3] Elimelech M, Phillip WA. The Future of Seawater Desalination: Energy, Technology, and the Environment. Science. 2011;333:712-7.

[4] Zhou L, Tan Y, Wang J, Xu W, Yuan Y, Cai W, et al. 3D self-assembly of aluminium nanoparticles for plasmon-enhanced solar desalination. Nat. Photonics. 2016;10:393-8.

[5] Tao P, Ni G, Song C, Shang W, Wu J, Zhu J, et al. Solar-driven interfacial evaporation. Nat. Energy. 2018;3:1031-41.

[6] Peng G, Sharshir SW, Wang Y, An M, Ma D, Zang J, et al. Potential and challenges of improving solar still by micro/nano-particles and porous materials - A review. J. Cleaner Prod. 2021;311:127432.

[7] Manokar AM, Winston DP, Mondol JD, Sathyamurthy R, Kabeel AE, Panchal H. Comparative study of an inclined solar panel basin solar still in passive and active mode. Sol. Energy. 2018;169:206-16.

[8] Morciano M, Fasano M, Bergamasco L, Albiero A, Lo Curzio M, Asinari P, et al. Sustainable freshwater production using passive membrane distillation and waste heat recovery from portable generator sets. Appl. Energy. 2020;258:114086.

[9] Shoeibi S, Rahbar N, Esfahlani AA, Kargarsharifabad H. Energy matrices, exergoeconomic and enviroeconomic analysis of air-cooled and water-cooled solar still: Experimental investigation and numerical simulation. Renewable Energy. 2021;171:227-44.

[10] Chen S, Zhao P, Xie G, Wei Y, Lyu Y, Zhang Y, et al. A floating solar still inspired by continuous root water intake. Desalination. 2021;512:115133.

[11] Panchal H, Sathyamurthy R, Kabeel AE, El-Agouz SA, Rufus D, Arunkumar T, et al. Annual performance analysis of adding different nanofluids in stepped solar still. J. Therm. Anal. Calorim. 2019;138:3175-82.

[12] Sharshir SW, Elkadeem MR, Meng A. Performance enhancement of pyramid solar distiller using nanofluid integrated with v-corrugated absorber and wick: An experimental study. Appl. Therm. Eng. 2020;168:114848.

[13] Peng G, Sharshir SW, Hu Z, Ji R, Ma J, Kabeel AE, et al. A compact flat solar still with high performance. Int. J. Heat Mass Transfer. 2021;311:127432.

[14] Sharshir SW, Elsheikh AH, Peng G, Yang N, El-Samadony MOA, Kabeel AE. Thermal performance and exergy analysis of solar stills – A review. Renewable Sustainable Energy Rev. 2017;73:521-44.

[15] Bait O. Direct and indirect solar–powered desalination processes loaded with nanoparticles: A review. Sustain. Energy Techn. 2020;37:100597.

[16] Xu Z, Zhang L, Zhao L, Li B, Bhatia B, Wang C, et al. Ultrahigh-efficiency desalination via a thermally-localized multistage solar still. Energy Environ. Sci. 2020;13:830-9.

[17] Li J, Wang X, Lin Z, Xu N, Li X, Liang J, et al. Over 10 kg m−2 h−1 Evaporation Rate Enabled by a 3D Interconnected Porous Carbon Foam. Joule. 2020;4:1-10.

[18] Sharshir SW, Algazzar AM, Elmaadawy KA, Kandeal AW, Elkadeem MR, Arunkumar T, et al. New hydrogel materials for improving solar water evaporation, desalination and wastewater treatment: A review. Desalination. 2020;491:114564.

[19] Jiang M, Shen Q, Zhang J, An S, Ma S, Tao P, et al. Bioinspired Temperature Regulation in Interfacial Evaporation. Adv. Funct. Mater. 2020:1910481.

[20] Ni G, Zandavi SH, Javid SM, Boriskina SV, Cooper TA, Chen G. A salt-rejecting floating solar still for low-cost desalination. Energy Environ. Sci. 2018;11:1510-9.

[21] Wang Z, Horseman T, Straub AP, Yip NY, Li D, Elimelech M, et al. Pathways and challenges for efficient solar-thermal desalination. Sci. Adv. 2019;5:eaax0763.

[22] Sharshir SW, Peng G, Wu L, Essa FA, Kabeel AE, Yang N. The effects of flake graphite nanoparticles, phase change material, and film cooling on the solar still performance. Appl. Energy. 2017;191:358-66.

[23] Parsa SM, Rahbar A, Koleini MH, Aberoumand S, Afrand M, Amidpour M. A renewable energy-driven thermoelectric-utilized solar still with external condenser loaded by silver/nanofluid for simultaneously water disinfection and desalination. Desalination. 2020;480:114354.

[24] Rabhi K, Nciri R, Nasri F, Ali C, Ben Bacha H. Experimental performance analysis of a modified single-basin single-slope solar still with pin fins absorber and condenser. Desalination. 2017;416:86-93.

[25] Sharshir SW, Elsheikh AH, Ellakany YM, Kandeal AW, Edreis EMA, Sathyamurthy R, et al. Improving the performance of solar still using different heat localization materials. Environ. Sci. Pollut. Res. 2020;27:12332-44.

[26] Shi J, Luo X, Liu Z, Fan J, Luo Z, Zhao C, et al. Efficient and antifouling interfacial solar desalination guided by a transient salt capacitance model. Cell Rep. Phys. Sci. 2021;2:100330.

[27] Taamneh Y, Taamneh MM. Performance of pyramid-shaped solar still: Experimental study. Desalination. 2012;291:65-8.

[28] Elango C, Gunasekaran N, Sampathkumar K. Thermal models of solar still—A comprehensive review. Renewable Sustainable Energy Rev. 2015;47:856-911.

[29] Dunkle R. Solar water distillation the roof type still and a multiple effect diffusion still. Int. Develop. Heat Transfer. 1961;5:895–902.

[30] Kumar S, Tiwari GN. Estimation of convective mass transfer in solar distillation systems. Sol. Energy. 1996;57:459-64.

[31] Zheng Hongfei, Zhang Xiaoyan ZJ, Yuyuan W. A group of improved heat and mass transfer correlations in solar stills. Energy Convers. Manage. 2002;43: 2469–78.

[32] Tsilingiris PT. The influence of binary mixture thermophysical properties in the analysis of heat and mass transfer processes in solar distillation systems. Solar Energy. 2007;81:1482–91.

[33] Peng G, Deng S, Sharshir SW, Ma D, Kabeel AE, Yang N. High efficient solar evaporation by airing multifunctional textile. Int. J. Heat Mass Transfer. 2020;147:118866.

[34] Tsilingiris PT. Review and critical comparative evaluation of moist air thermophysical properties at the temperature range between 0 and 100 °C for Engineering Calculations. Renewable Sustainable Energy Rev. 2018;83:50-63.

# Supporting information

## Improving the mass transfer rate and energy efficiency of solar still by enhancing the inner air circulation

Guilong Peng, Zhenwei Xu, Jiajun Ji, Senshan Sun, Nuo Yang*

State Key Laboratory of Coal Combustion, Huazhong University of Science and Technology, Wuhan 430074, China

## Note S1 Verification of the thermostat cover

Due to the instability of the ambient temperature, the thermostat cover is used to control the ambient temperature near the glass cover of the solar still. To verify the effectiveness of the thermostat cover, the results of the real ambient and simulated ambient is compared. As shown in Fig. S1, the energy efficiency of the real ambient and the simulated ambient is very close, which shows the validity of using the thermostat cover.

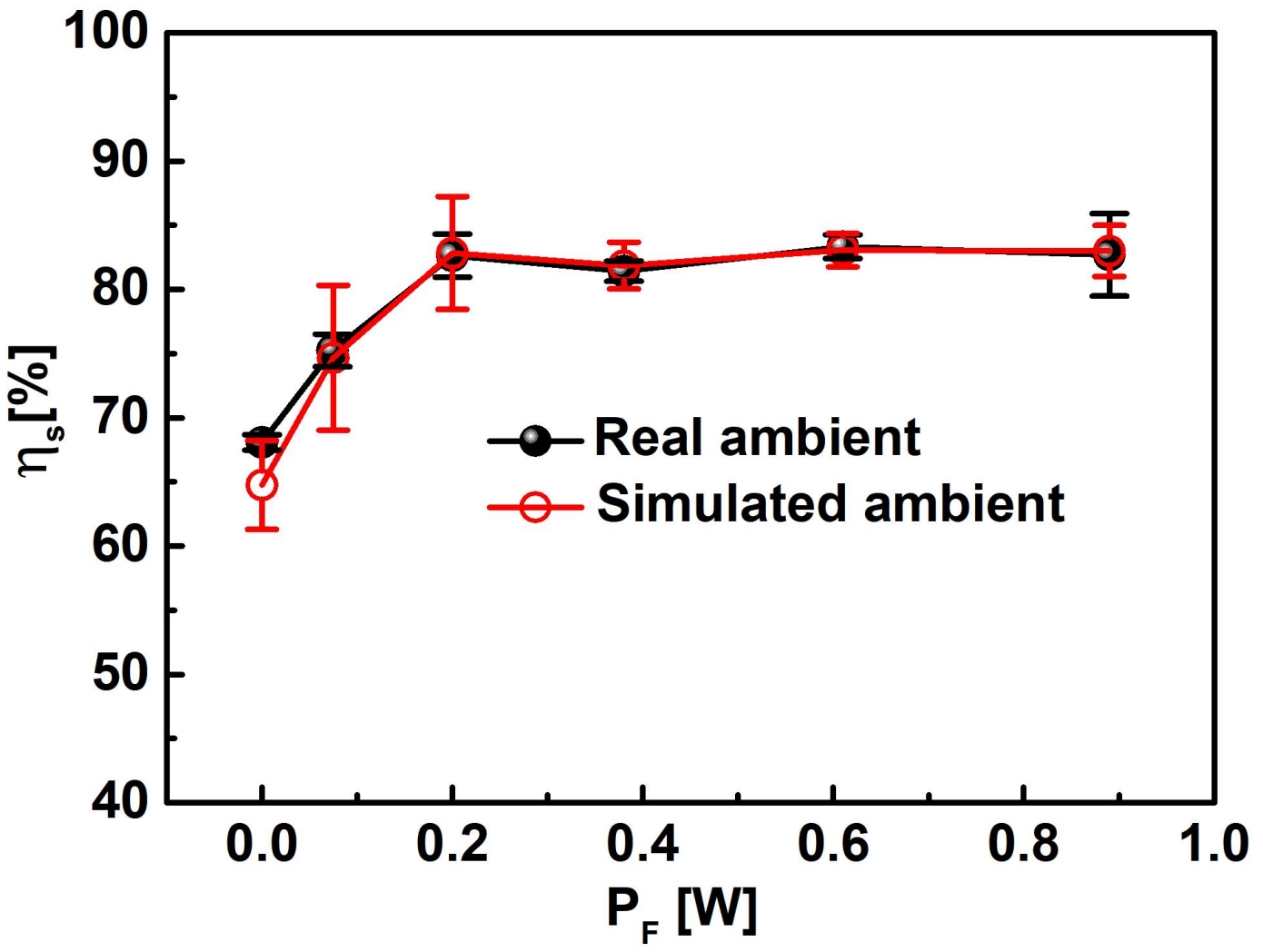

Figure. S1 The energy efficiency of the solar still. The input power density is fixed at 500 W/m$^2$. The real ambient indicates that the solar still is working without the thermostat cover, and the real ambient temperature is almost stable at 30°C. The simulated ambient indicates that the solar still is working with thermostat cover, and the temperature of the thermostat cover is 30°C.

**Note S2 The measurement of temperatures**

To measure the temperature of the glass cover and the water, six thermal couples are used in the solar still as shown in Fig. S2(a). In the main text, the glass temperature is regarded as the average of $T_{g1}$, $T_{g2}$ and $T_{g3}$. The water temperature is regarded as the average of $T_{w1}$, $T_{w2}$ and $T_{w3}$. The temperatures of the water and the glass cover under a typical condition is shown in Fig. S2(b). With the fan operated, the water temperature and the glass temperature at the edge decrease a lot at first, which shows that the heat and mass transfer is concentrated in the center of the solar still. The temperature of foam and ambient are also measured for analysis.

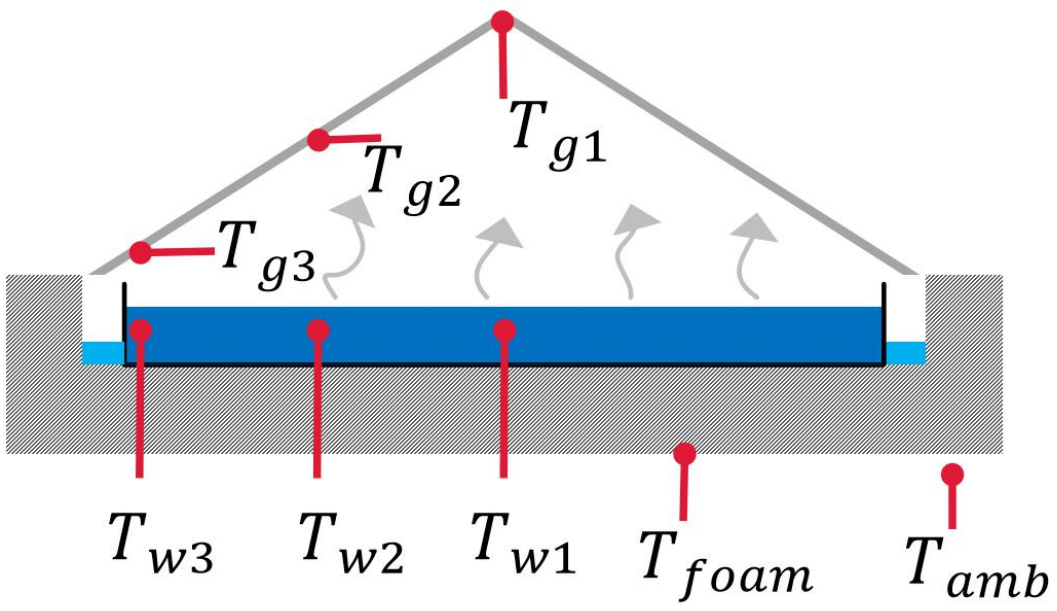

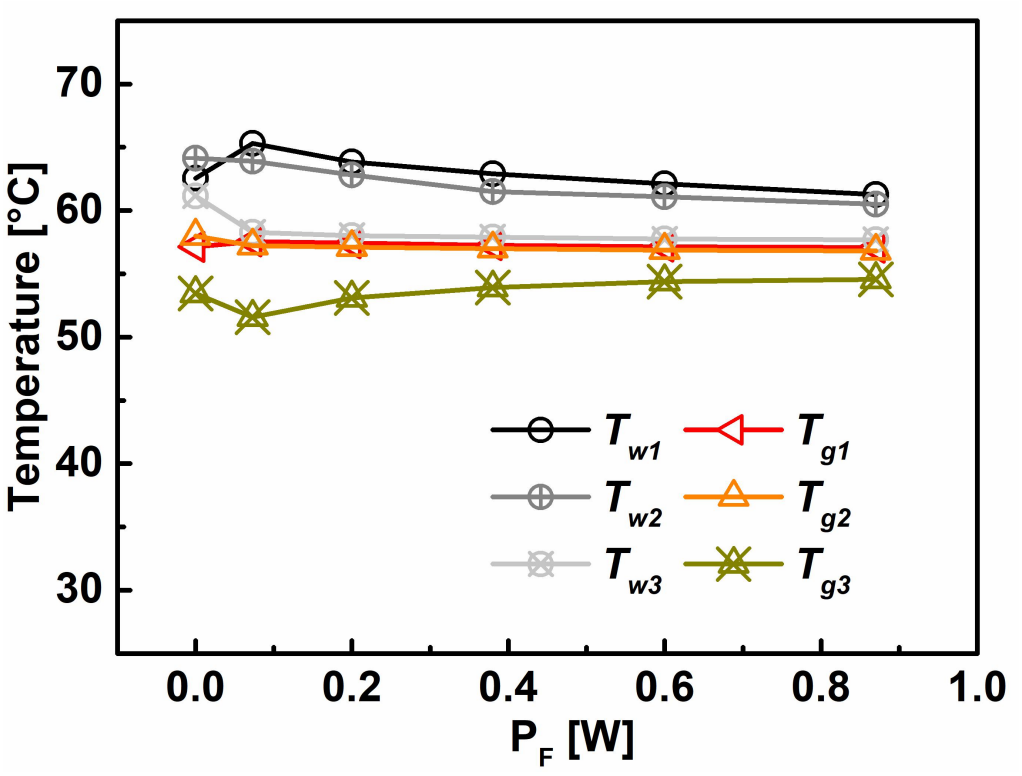

Figure. S2 (a)The schematic diagram of the thermal couples in the solar still. (b) The temperatures of the water and the glass cover at $I$ = 500 W/m$^2$ and $T_{amb}$ = 30°C.

## Note S3 Effect of the ambient temperature

The experimental results under different ambient temperatures also agree well with the theoretical results (Fig. S3). With the increases of the ambient temperature, the energy efficiency increases slightly. It indicates that under the same input power density, the energy efficiency in hotter regions or seasons will be higher.

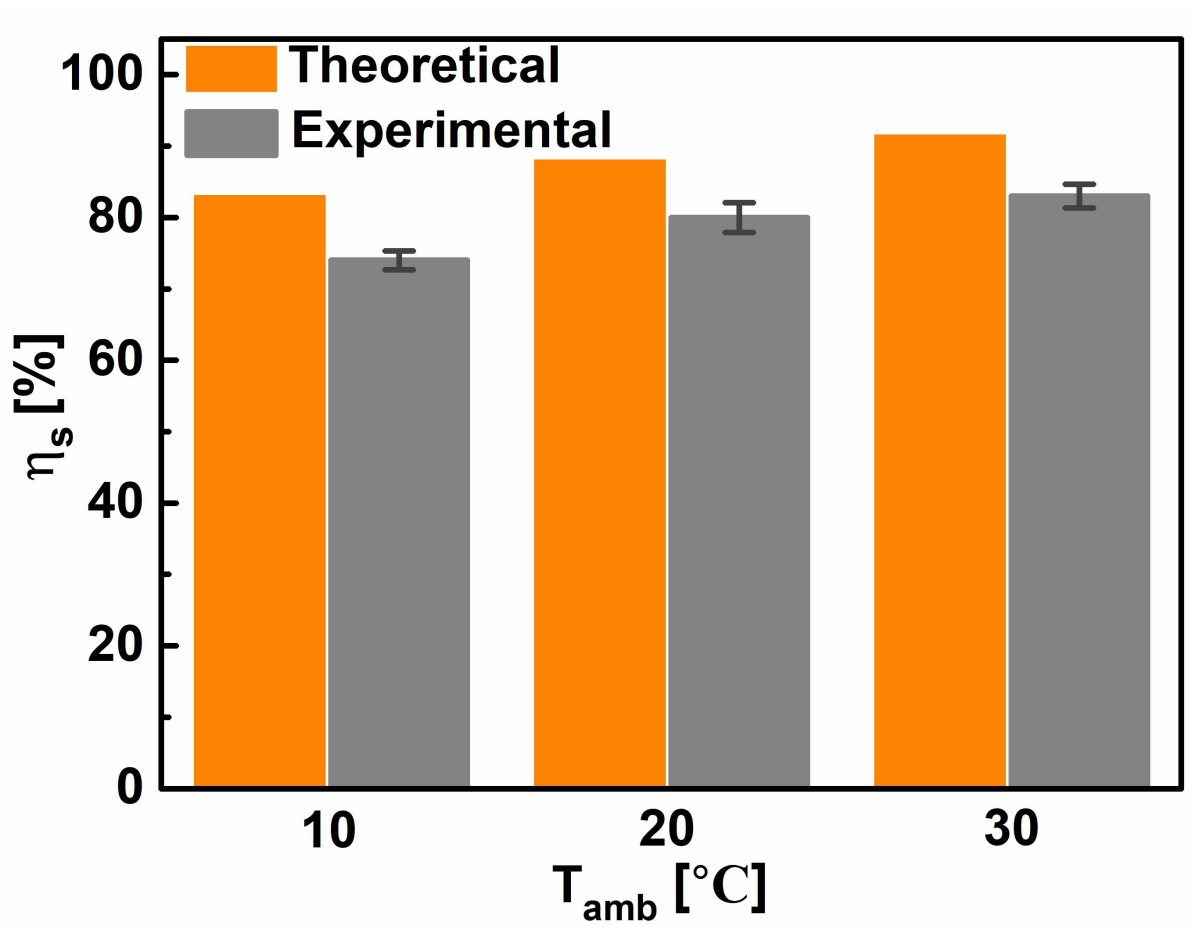

Figure. S3 The maximum energy efficiency of the pyramid SS under different ambient temperatures. The input power density is fixed at 500 W/m$^2$.